\documentclass[preprint,12pt]{aastex}

\shortauthors{Winn et al.~2006}
\shorttitle{Two Transits of TR-111}

\begin{document}

%
\def\ltsima{$\; \buildrel < \over \sim \;$}
\def\lsim{\lower.5ex\hbox{\ltsima}}
\def\gtsima{$\; \buildrel > \over \sim \;$}
\def\gsim{\lower.5ex\hbox{\gtsima}}
\def\rs{$R_S = 0.831 \pm 0.031$~$R_\odot$}
\def\rp{$R_P = 1.067 \pm 0.054$~$R_{\rm Jup}$}
                                                                                          
%

\bibliographystyle{apj}

\title{
The Transit Light Curve (TLC) Project.\\
II.~Two Transits of the Exoplanet OGLE-TR-111b
}

\author{
Joshua N.\ Winn\altaffilmark{1},
Matthew J.\ Holman\altaffilmark{2},
Cesar I.\ Fuentes\altaffilmark{2}
}

\altaffiltext{1}{Department of Physics, and Kavli Institute for
  Astrophysics and Space Research, Massachusetts Institute of
  Technology, Cambridge, MA 02139}

\altaffiltext{2}{Harvard-Smithsonian Center for Astrophysics, 60
  Garden Street, Cambridge, MA 02138}

\begin{abstract}

  As part of our ongoing effort to measure exoplanet sizes and transit
  times with greater accuracy, we present $I$ band observations of two
  transits of OGLE-TR-111b. The photometry has an accuracy of
  0.15-0.20\% and a cadence of 1--2~minutes. We derive a planetary
  radius of \rp~and a stellar radius of \rs. The uncertainties are
  dominated by errors in the photometry, rather than by systematic
  errors arising from uncertainties in the limb darkening function or
  the stellar mass. Both the stellar radius and the planetary radius
  are in agreement with theoretical expectations. The transit times
  are accurate to within 30 seconds, and allow us to refine the
  estimate of the mean orbital period: $4.0144479\pm 0.0000041$~days.

\end{abstract}


\section{Introduction}

The aim of the Transit Light Curve (TLC) project is to gather accurate
photometry during the transits of exoplanets across the disks of their
parent stars. The immediate scientific harvest is the improved
accuracy with which the basic system parameters are
known. High-accuracy, high-cadence transit photometry allows for the
determination of the stellar radius, the planetary radius, and the
orbital inclination, for an assumed value of the stellar mass (see,
e.g., Seager \& Mall\'{e}n-Ornelas~2003 for a discussion of the
theory, and Brown et al.~2001 for the most famous example of a transit
light curve). These parameters are interesting in their own right, and
are important in the interpretation of other measurements such as
reflected light (Rowe et al.~2006), thermal emission (Charbonneau et
al.~2005, Deming et al.~2005), atmospheric absorption (Charbonneau et
al.~2002, Vidal-Madjar et al.~2003), and the Rossiter-McLaughlin
effect (Queloz et al.~2000, Winn et al.~2005, Wolf et al.~2006).

In the longer term, repeated observations of transits will reduce the
uncertainties in the system parameters (and especially the orbital
period) still further. More interestingly, it may be possible to
detect additional transiting objects, satellites, rings, or even
reflected light, by combining the data from many individual light
curves. In addition, the existence of hitherto-undetected planets and
satellites might be betrayed by periodic patterns in the measured
times of mid-transit or variations in the orbital inclination
(Miralda-Escud\'{e}~2002, Holman \& Murray 2005, Agol et al.~2005).

The TLC project is currently in its initial phase, in which we are
observing all of the known transiting exoplanets, refining the
estimates of each system's parameters and assessing the feasibility of
continued long-term monitoring. We have previously reported on
observations of the exoplanet XO-1b (Holman et al.~2006). In this
paper, we present TLC results for OGLE-TR-111b.

The periodic 2\% dimming events of the star OGLE-TR-111 were
discovered by Udalski et al.~(2002) in a survey for transiting planets
within a rich star field in Carina. Spectroscopic follow-up by Pont et
al.~(2004) revealed a periodic Doppler shift, confirming that the
dimming events were caused by the transits of a Jovian planet. Santos
et al.~(2006) obtained optical spectra with a higher signal-to-noise
ratio to study the properties of the host star, which is an early K
dwarf with roughly solar metallicity.  The orbital period is just over
4~days, which is the longest period among the 5 exoplanets identified
in the OGLE survey, but which is typical of the periods of the ``hot
Jupiters'' that have been discovered in abundance in radial-velocity
surveys.  For this reason, Pont et al.~(2004) referred to OGLE-TR-111
as the ``missing link'' between the OGLE survey and the
radial-velocity surveys.  Initially, it was thought that the two
surveys were yielding discrepant results because of the shorter
periods of the OGLE objects, but this is now understood as a selection
effect (Pont et al.~2005, Gaudi et al.~2005, Gould et al.~2006).

This paper is organized as follows. We describe the observations in
the next section, and the photometric procedure in \S~3. In \S~4, we
describe the techniques we used to estimate the physical and orbital
parameters. In \S~5 we give the results, and the final section is a
brief summary.

\section{Observations}

We observed two transits of OGLE-TR-111 (on UT~2006~Feb~21 and Mar~5)
corresponding to epochs $E=363$ and 366 of an updated ephemeris based
on OGLE data that was provided by A.~Udalski (2005, private
communication):
\begin{equation}
T_c(E) = 2,452,330.46228~{\mathrm{[HJD]}} + E\times(4.014442~{\mathrm{days}}).
\end{equation}

We used the Inamori Magellan Areal Camera and Spectrograph (IMACS) on
the 6.5m Baade (Magellan~I) telescope at Las Campanas Observatory, in
Chile. IMACS has two cameras differing in focal length. We used the
longer $f/4.3$ camera because we preferred the smaller pixel scale,
and the field of view was still large enough to encompass multitudes
of comparison stars. Photometry is improved with small pixels not only
because of the better spatial sampling of the PSF, but also because
spreading the starlight over many pixels averages down the
pixel-to-pixel sensitivity variations and increases the maximum
exposure time due to saturation. The IMACS detector is a mosaic of
eight 2k~$\times$~4k SITe back-illuminated and thinned CCDs with
15~$\mu$m pixels, giving a pixel scale of $0\farcs111$ and a field of
view of $15\farcm4$. To reduce the readout time, we read only
one-third of each chip, corresponding to the central $15\farcm4 \times
5\farcm1$ of the mosaic. The readout time was approximately 45~s and
the readout noise was about 5~e$^{-}$. We observed through the CTIO
$I$~band filter, the reddest broad-band filter in routine use on
IMACS, in order to minimize the effect of color-dependent atmospheric
extinction on the relative photometry, and to minimize the effect of
limb-darkening on the transit light curve. On each of the two nights,
we observed OGLE-TR-111 for approximately 6~hr bracketing the
predicted midpoint of the transit. We also obtained dome flat
exposures and zero-second (bias) exposures at the beginning of each
night.

On the night of UT~2006~Feb~21, we observed under clear skies, through
an airmass ranging from 1.2 to 1.6. The seeing was generally good but
variable, from $0\farcs5$ to $1\farcs0$. We used an exposure time of
60~s. Although the sky conditions were excellent, three factors
degraded the photometry to some degree. First, although we attempted
to keep the image registration constant throughout the night (thereby
consistently detecting the light from each star on the same set of
pixels), this was not possible due to occasional failures of the guide
probe control software. The changes in registration had a noticeable
effect on the relative photometry, as described further in
\S~3. Second, the diffraction spike from a nearby bright star swept
through the position of OGLE-TR-111 on two occasions, with noticeable
effects on the photometry. Third, the atmospheric dispersion corrector
(ADC) was not functioning correctly, causing color-dependent effects
in the stellar images.

On the night of UT~2006~Mar~05, the skies were also clear, and the
seeing was more consistent at approximately $0\farcs9$ all night. We
used a shorter exposure time of 30~s.  As in February, there were
occasional crashes of the guiding software, and a diffraction spike
swept through the position of OGLE-TR-111.  However, on this night the
ADC was functioning properly.

\section{Data Reduction}

We used standard IRAF\footnote{ The Image Reduction and Analysis
  Facility (IRAF) is distributed by the National Optical Astronomy
  Observatories, which are operated by the Association of Universities
  for Research in Astronomy, Inc., under cooperative agreement with
  the National Science Foundation. } procedures for the overscan
correction, trimming, bias subtraction, and flat-field
division. Because the images were too crowded for aperture photometry,
we used the method of image subtraction as implemented by Alard \&
Lupton~(1998) and Alard~(2000). Specifically, we used version 2.2 of
the ISIS image subtraction package that was written and kindly made
public by C.~Alard. In this method, all of the images from a given
night are registered to a common pixel frame, and a reference image is
created by combining a subset ($\approx$10\%) of the images with the
best seeing. For each individual image, a convolution kernel is
determined that brings the image into best agreement with the
reference image. Then the difference is computed between the
appropriately convolved image and the reference image. The advantage
of this method is that photometry is simplified on the difference
images, because most stars are not variable stars and thus the complex
and crowded background is eliminated. It is still necessary to compute
the flux of the variable stars on the reference image (taking into
account any neighboring stars) but this need only be done once, and
the task is facilitated by the good spatial resolution and high
signal-to-noise ratio of the reference image. Thus, the measurement of
the relative flux $f(t)$ takes the form
\begin{equation}
f(t) = 1 + \frac{\Delta f(t)}{f_{\rm ref}},
\end{equation}
where $f_{\rm ref}$ is measured on the reference image, and $\Delta
f(t)$ is measured on the difference images.

We performed photometry of OGLE-TR-111 along with 18 other stars of
comparable brightness for quality control. We tried two different
methods for performing the photometry on the difference images:
IRAF-based aperture photometry and ISIS-based profile-fitting
photometry. For the March data, superior results (in the sense of a
smaller standard deviation in the light curves of the comparison
stars) were obtained with profile-fitting photometry, whereas for the
February data, superior results were obtained with aperture
photometry. We suspect that the reason for the difference is that the
February data were taken under more variable seeing and without proper
atmospheric dispersion correction. Because of these differing
conditions, we adopted the aperture results for February, and the
profile-fitting results for March, rather than requiring the same
procedure to be used on all of the data. To determine $f_{\rm ref}$
for each star, we performed profile-fitting photometry on the
reference image for each night.

The uncertainty in each data point arises from two sources: the
uncertainty in the difference flux $\Delta f$, and the uncertainty in
the reference flux $f_{\rm ref}$. We estimated the uncertainty in the
difference flux based on Poisson statistics. We estimated the
uncertainty in the reference flux based not only on Poisson
statistics, but also by the spread in the values obtained when using
different choices for the stars used to determine the
point-spread-function (PSF) and other parameters relating to the
profile photometry. This latter source of systematic error was 1.5\%
for the February data and 1\% for the March data, which dominated the
Poisson error in both cases. However, adjustments in $f_{\rm ref}$
affect all of the points from a given night in the same way; the net
effect is a small modification of the transit depth. For example, for
OGLE-TR-111 the transit depth is approximately 2\%. The effect of
increasing $f_{\rm ref}$ by 1\% is to decrease the transit depth by
$(0.02 \times 0.01)$ or $2\times 10^{-4}$. We discuss this systematic
error further in \S~5.

Abrupt jumps in the photometry by $\sim$0.5\% were evident when the
diffraction spike from a nearby star intruded on the position of
OGLE-TR-111, and when the telescope pointing changed (presumably due
to flat fielding errors). We discarded the data that was affected by
the diffraction spike. For the February data, the pointing changed
approximately 15~minutes prior to ingress, and again just after
egress; in our subsequent analysis we used only the data obtained
between those times. Most of the March data was taken with a common
pointing, except for an interval of 30~minutes after egress, which
occurred after a guider failure and before the pointing could be
restored to its former value. Those 30 minutes of data were not
considered further.

\begin{figure}[p]
\epsscale{0.9}
\plotone{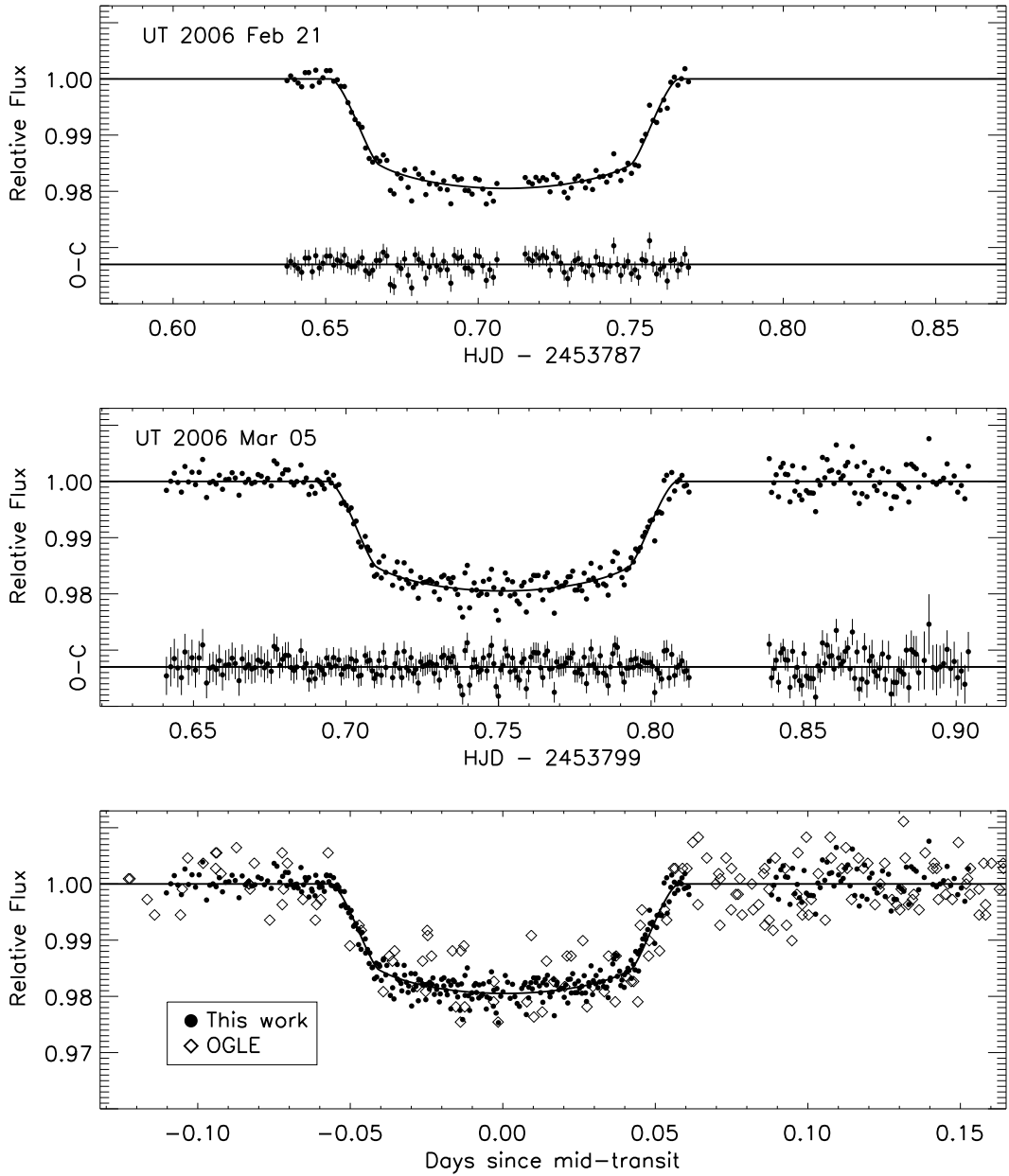}
\caption{Relative $I$ band photometry of OGLE-TR-111. The best-fitting
model is shown as a solid line. The residuals
(observed~$-$~calculated) and the rescaled $1~\sigma$ error bars are
also shown. The residuals have zero mean but are offset by a constant
flux to appear beneath each light curve, for clarity.  The
root-mean-squared residuals are 0.15\% and 0.2\% for the
February and the March data, respectively.  The lowest panel shows the
composite light curve and also a composite light curve based on the
OGLE survey data.
\label{fig:lc}}
\end{figure}

Although the image subtraction method removes the first-order effects
of extinction by scaling all of the images to a common flux level
before subtraction, residual color-dependent effects are not removed.
Stars of different colors are extinguished by different amounts
through a given airmass.  For this reason, we applied a residual
extinction correction to the data.  The correction function was
determined as part of the model-fitting procedure and will be
described in the next section.  The final photometry is given in
Table~1, and is plotted in Fig.~\ref{fig:lc}. In the bottom panel of
Fig.~1, our composite light curve is compared to that of the OGLE
survey data. The uncertainties given in Table~1 are the uncertainties
in the difference fluxes, after multiplying by a factor specific to
each night such that $\chi^2/N_{\rm DOF} = 1$ for the best-fitting
model. (Our intention was not to test the model, but rather to
determine the appropriate relative weights for the data points.) The
scaling factors were 1.32 for the February data and 1.01 for the March
data.

\section{The Model}

Our model for the system is based on a star (with mass $M_S$ and
radius $R_S$) and a planet (with mass $M_P$ and radius $R_P$) in a
circular orbit\footnote{A circular orbit is a reasonable simplifying
  assumption because it is expected that there has been sufficient
  time for tides to have damped out any initial eccentricity, in the
  absence of a third body (see, e.g., Rasio et al.~1996, Trilling et
  al.~2000, Dobbs-Dixon et al.~2004).} with period $P$ and inclination
$i$ relative to the sky plane. We define the coordinate system such
that $0\arcdeg \leq i\leq 90\arcdeg$. We allow each transit to have an
independent value of $T_c$, the transit midpoint, rather than forcing
them to be separated by an integral number of orbital periods. This is
because we seek to measure or bound any timing anomalies that may
indicate the presence of moons or additional planets in the
system. Thus, the period $P$ is relevant to the model only through the
connection between the total mass and the orbital semimajor axis. We
fix $P=4.01444$~days, the value kindly provided by A.~Udalski (2005,
private communication) based on several seasons of OGLE data.

The stellar mass cannot be determined from transit photometry alone.
Furthermore, the values of $R_S$ and $R_P$ that are inferred from the
photometry are covariant with the stellar mass; for a fixed period
$P$, the photometric transit depends almost exactly on the
combinations $R_S/M_S^{1/3}$ and $R_P/M_S^{1/3}$. Our approach was to
fix $M_S = 0.81$~$M_\odot$, the value reported by by Santos et
al.~(2006) based on an analysis of the stellar spectrum (i.e., the
observed effective temperature, surface gravity, and metallicity were
compared to theoretical H-R diagrams). We then use the scaling
relations $R_P \propto M_S^{1/3}$ and $R_S \propto M_S^{1/3}$ to
estimate the systematic error due to the uncertainty in $M_S$. The
planetary mass $M_P$ is nearly irrelevant to the model (except for its
minuscule effect on the relation between the orbital period and the
semimajor axis), but for completeness we use the value
$M_P=0.52$~$M_{\rm Jup}$ reported by Santos et al.~(2006).

To calculate the relative flux as a function of the projected
separation of the planet and the star, we assumed the limb darkening
law to be linear,
\begin{equation}
\frac{I(\mu)}{I(1)} = 1 - u(1-\mu),
\end{equation}
where $I$ is the intensity, and $\mu$ is the cosine of the angle
between the line of sight and the normal to the stellar surface.
Adding a parameter to the limb darkening law by making it a quadratic
function of $(1-\mu)$ did not significantly improve the fit, and hence
is not well justified by the data alone. We employed the analytic
formulas of Mandel \& Agol~(2002) to compute the integral of the
intensity over the exposed portion of the stellar disk. The limb
darkening parameter $u$ was a variable in the model, but we applied an
${\it a~priori}$ constraint to require a reasonable level of agreement
with theoretical expectations for limb darkening, as described below.
Each transit is also described with two additional parameters: the
out-of-transit flux $f_{\rm oot}$, and a residual extinction
coefficient $k$. The latter is defined such that the observed flux is
proportional to $\exp(-kz)$.

In total, there are 10 adjustable parameters describing 386
photometric data points. The parameters are $R_S$, $R_P$, and $i$; the
two values of $T_c$; the limb-darkening parameter $u$; and the values
of $f_{\rm oot}$ and $k$ for each transit.  Our goodness-of-fit
parameter is
\begin{equation}
\chi^2 = \sum_{j=1}^{386}
\left[
\frac{f_j({\mathrm{obs}}) - f_j({\mathrm{calc}})}{\sigma_j}
\right]^2 +
\left[
\frac{u - u_{\rm th}}{\sigma_u}
\right]^2
\end{equation}
where $f_j$(obs) is the flux observed at time $j$, $\sigma_j$ is the
corresponding uncertainty, and $f_j$(calc) is the calculated value.
The last term is the {\it a priori} constraint on the limb darkening
parameter. The theoretical value $u_{\rm th} =0.597$ comes from fits
by Claret~(2000) to an ATLAS plane-parallel stellar atmosphere model
of R.~Kurucz, for a star with $T_{\rm eff}=5000$~K, $\log g =
4.5$~(cgs), and $[{\rm Fe}/{\rm H}] = 0.2$.  We set $\sigma_u=0.081$,
corresponding to the requirement that the limb-to-center intensity
ratio ($1-u$) should be within about 20\% of the calculated value. (We
also investigated the effects of tightening, modifying and dropping
this {\it a priori} constraint, as discussed in the next section.) As
noted in \S~3, we took the uncertainties $\sigma_j$ to be the
calculated uncertainties after multiplication by a factor specific to
each night, such that $\chi^2/N_{\rm DOF} = 1$ when each night's data
was fitted individually.

We began by finding the values of the parameters that minimize
$\chi^2$, using the venerable AMOEBA algorithm~(Press et al.~1992, p.\
408). Then we estimated the {\it a posteriori} joint probability
distribution for the parameter values using a Markov Chain Monte Carlo
(MCMC) technique (for a brief introduction, consult appendix A of
Tegmark et al.~2004). In this method, a chain of points in parameter
space is generated from an initial point by iterating a jump function,
which in our case was the addition of a Gaussian random number to each
parameter value. If the new point has a lower $\chi^2$ than the
previous point, the jump is executed; if not, the jump is only
executed with probability $\exp(-\Delta\chi^2/2)$. We set the typical
sizes of the random perturbations such that $\sim$20\% of jumps are
executed. We created 10 independent chains, each with 500,000 points,
starting from random initial positions.  The first 100,000 points were
not used, to minimize the effect of the initial condition.  The
correlation lengths were $\sim$$10^3$ steps for the highly covariant
parameters $R_P$, $R_S$, and $b$, and a few hundred steps for the
other parameters.  The Gelman \& Rubin~(1992) $R$ statistic was
within 0.1\% of unity for each parameter, a sign of good mixing and
convergence.

\section{Results}

The model that minimizes $\chi^2$ is plotted as a solid line in
Fig.~\ref{fig:lc}. The optimized residual extinction correction has
been applied to the data that are plotted in Fig.~\ref{fig:lc}, and to
the data that are given in Table~1. The differences between the
observed fluxes and the model fluxes are also shown beneath each light
curve.

\begin{figure}[p]
\epsscale{1.0}
\plotone{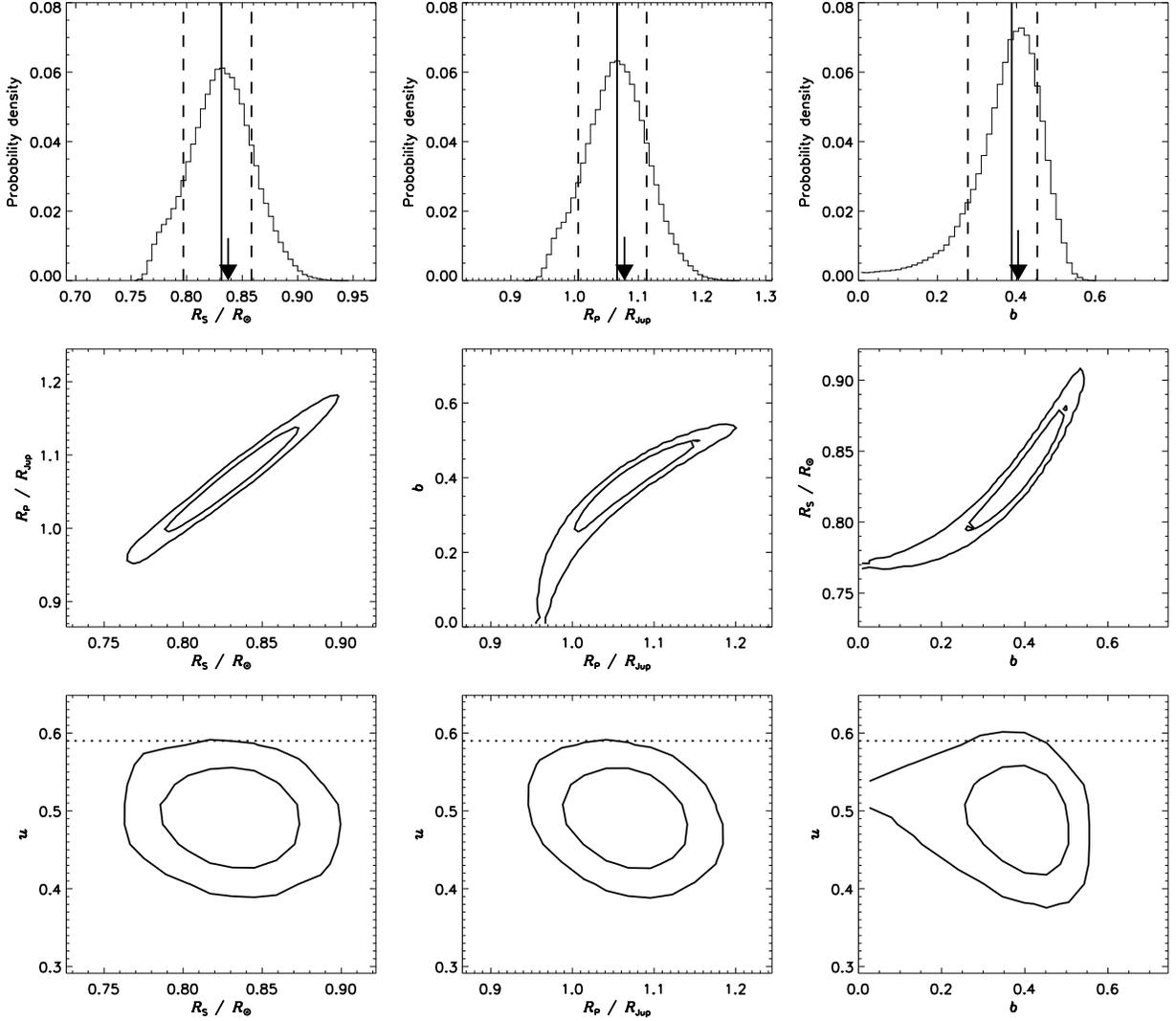}
\caption{ {\bf Top row.} Probability distributions for the stellar
  radius $R_S$, planetary radius $R_P$, and impact parameter $b\equiv
  a\cos i/R_S$, based on the MCMC simulations. The arrows mark the
  values of the parameters that minimize $\chi^2$. A solid line marks
  the median of each distribution, and the dashed lines mark the 68\%
  confidence limits.  {\bf Middle and bottom rows.}  Joint probability
  distributions of those parameters with the strongest correlations.
  The contours are isoprobability contours enclosing 68\% and 95\% of
  the points in the Markov chains. The dots mark the values
  of the parameters that minimize $\chi^2$. The dotted lines indicate
  the value of the limb darkening parameter calculated by Claret~(2000)
  based on an ATLAS model of a star with $T_{\rm eff}=5000$~K, $\log
  g=4.5$~(cgs), $[{\rm Fe}/{\rm H}]=0.2$, and $\xi_t=2$~km~s$^{-1}$.
\label{fig:err}}
\end{figure}

Table~2 gives the estimated values and uncertainties for each
parameter based on the MCMC analysis. It also includes some useful
derived quantities: the impact parameter $b= a \cos i / R_S$ (where
$a$ is the semimajor axis); the time between first and last contact
($t_{\rm IV} - t_{\rm I}$); and the time between first and second
contact ($t_{\rm II} - t_{\rm I}$). Fig.~\ref{fig:err} shows the
estimated {\it a posteriori} probability distributions for the
especially interesting parameters $R_S$, $R_P$ and $b$, along with
some of the two-dimensional correlations involving those
parameters. Although the distributions shown in Fig.~\ref{fig:err} are
somewhat asymmetric about the median, Table~2 reports only the median
$p_{\rm med}$ and a single number $\sigma_p$ characterizing the width
of the distribution. The value of $\sigma_p$ is the average of
$|p_{\rm med}-p_{\rm hi}|$ and $|p_{\rm med} - p_{\rm lo}|$, where
$p_{\rm lo}$ and $p_{\rm hi}$ are the lower and upper 68\% confidence
limits. We refer to $\sigma_p$ as the ``statistical error'' to
distinguish it from the sources of systematic error discussed below.

There are several sources of systematic error that are not taken into
account by the MCMC analysis. The first is the systematic error that
was already discussed in \S~4: the covariance between the stellar mass
$M_S$ and both of the parameters $R_P$ and $R_S$. For a fixed value of
$P$, the photometric signal depends on the combinations
$R_P/M_S^{1/3}$ and $R_S/M_S^{1/3}$. A value for $M_S$ must be chosen
on other grounds.  We adopted the value $M_S=0.81$~$M_\odot$, based on
the most recent analysis of the spectrum of OGLE-TR-111 by Santos et
al.~(2006).  Based on their measurements of the equivalent widths of
38 iron absorption lines and the wings of the H$\alpha$ absorption
profile, those investigators concluded that the uncertainty in $M_S$
was only 2.5\%. The corresponding fractional error in $R_P$ and $R_S$
due to the covariance is only 0.8\%, which is negligible when compared
to the statistical errors of 4\% and 5\%, respectively.  Another way
to state this result is that the uncertainty in the stellar mass would
need to be $\gsim$10\% (i.e., four times larger than the uncertainty
quoted by Santos et al.~2006) for the resulting systematic error to be
comparable to the statistical error.

A second source of systematic error is the bias due to an incorrect
choice of limb darkening function. Neither the appropriate functional
form of the limb darkening function, nor the value of the limb
darkening coefficient, is known with certainty. One can calculate the
limb darkening function based on stellar atmosphere models, but the
uncertainties in both the stellar parameters and in the atmosphere
models make it hard to quantify the uncertainty in the results. We
attempted to account for this uncertainty with the {\it a priori}
constraint on $u$ that was described in \S~4; here, we describe the
results of some tests of the sensitivity of our results on the
treatment of limb darkening. For brevity, we describe only the
variation in $R_P$ under different assumptions, because we have found
that this parameter shows the greatest sensitivity to changes in the
assumed limb-darkening law. If $u$ is held fixed at the Claret~(2000)
value of 0.597, then $R_P$ is decreased by 1.3\% relative to the value
given in Table~1. If $u$ is allowed to vary with no constraint, $R_P$
rises by 0.8\%.  Using a quadratic limb-darkening law instead of a
linear law, $R_P$ decreases by 2.5\% if the coefficients are fixed at
the Claret~(2000) values, and by 0.5\% if they vary freely. The
four-parameter ``nonlinear'' law gives essentially the same results as
the quadratic law, and the PHOENIX-based coefficients give essentially
the same results as the ATLAS coefficients. We conclude that the
systematic error in $R_P$ due to the choice of limb darkening law is a
few per cent at most, which is smaller than the statistical error.

A third source of systematic error is in the measurement of the flux
of OGLE-TR-111 on the reference image of the image-subtraction
photometric procedure (see \S~3). For example, if the reference flux
$f_{\rm ref}$ is erroneously large, then the transit depth (and the
inferred value of $R_P/R_S$) will be erroneously small.  We assessed
the size of this effect by re-fitting the data after adjusting the
value of $f_{\rm ref}$ for the March data (which dominates $\chi^2$)
upward or downwards by the estimated error. The results for $R_P$ and
$R_S$ are altered by $\lsim$0.5\%, a change that can be neglected in
light of the larger errors determined previously.

We believe that these three effects are the largest sources of
systematic error, and we have shown that all of them are smaller than
the statistical errors in the photometry.  Therefore, unless the
uncertainty in $M_S$ has been grossly underestimated by Santos et
al.~(2006), we conclude that there is considerable scope for
improvement in the system parameters through additional photometry.

Our value of \rp\ for the planetary radius is in agreement with
previous estimates. Pont et al.~(2004) found $R_P =
1.00^{+0.13}_{-0.06}$~$R_{\rm Jup}$ based on the OGLE photometry, and
Santos et al.~(2006) refined this value to $R_P = 0.97\pm
0.06$~$R_{\rm Jup}$. These values, in turn, have already shown to be
in broad agreement with theoretical expectations for ``hot Jupiters''
(see, e.g., Baraffe et al.~2005). It might seem surprising that our
estimate is hardly more precise than that of Santos et al.~(2006),
despite our superior photometry. However, the comparison is
misleading. The time sampling and signal-to-noise ratio of the OGLE
photometry were insufficient for reliable measurements of the ingress
and egress durations. Consequently, it was necessary for previous
transit light-curve fitters to adopt an {\it a priori} value of $R_S$
in addition to $M_S$. We have been able to derive $R_S$ from the
photometry, subject only to the weak covariance with the assumed value
of $M_S$. The agreement between our result \rs\ and the spectroscopic
estimate of $0.83\pm 0.02$~$R_\odot$ is therefore a new and important
cross-check on the system parameters.

The uncertainties in the transit times are about 30 seconds, and these
uncertainties are not correlated with those of any of the other model
parameters except for the residual airmass correction. The interval
between the two transits was $12.0428(5)$~days, where the number in
parentheses is the 1$\sigma$ error in the last digit. This corresponds
to an ``instantaneous period'' of $4.0143(2)$~days. We refined the
precision of the transit ephemeris through a simultaneous fit to all
of the OGLE photometry and our own photometry, assuming the period to
be uniform. Only the orbital period $P$ and one particular time of
transit $T_c$ were allowed to vary; all of the rest of the parameters
were held fixed at the values given in Table~1. The refined ephemeris
is
\begin{eqnarray}
T_c & = & 2,453,799.7516 \pm 0.0002~{\mathrm{[HJD]}} \nonumber \\
P   & = & 4.0144479\pm 0.0000041~{\mathrm{days}}.
\end{eqnarray}

\section{Summary}

Through observations of two closely spaced transits of the exoplanet
OGLE-TR-111b, we have improved upon the estimates of the system
parameters. The improvement comes not only from an overall increase in
the signal-to-noise ratio, but also from the ability to resolve the
ingress and egress and thereby determine the stellar radius
photometrically. Our results confirm the previous estimates of the
stellar and planetary radii, and are subject to a smaller systematic
error. We have also provided a more precise transit ephemeris. All of
these results will help with future observations and interpretations
of this system.

\acknowledgments We have benefited from helpful consultations with
D.~Sasselov on limb darkening, G.~Torres on stellar mass
determination, and S.~Gaudi on parameter estimation. We thank
A.~Udalski for providing an updated OGLE ephemeris and A.~Roussanova
for proofreading the manuscript.

\begin{deluxetable}{lcccc}
\tabletypesize{\normalsize}
\tablecaption{Photometry of OGLE-TR-111\label{tbl:photometry}}
\tablewidth{0pt}

\tablehead{
\colhead{HJD} & \colhead{Relative flux} & \colhead{Uncertainty}
}

\startdata
$ 2453787.637339 $  &  $1.00005$  &  $0.00158$  \\
\enddata 

\tablecomments{The time stamps represent the Heliocentric Julian Date
  at the time of mid-exposure. The uncertainty estimates are based on
  the procedures described in \S~2. We intend for this table to appear
  in entirety in the electronic version of the journal. A portion is
  shown here to illustrate its format. The data are also available
  from the authors upon request.}

\end{deluxetable}

\begin{deluxetable}{ccc}
\tabletypesize{\small}
\tablecaption{System Parameters of OGLE-TR-111\label{tbl:params}}
\tablewidth{0pt}

\tablehead{
\colhead{Parameter} & \colhead{Value} & \colhead{Uncertainty}
}
\startdata
                                               $R_S/R_\odot$& $          0.831$ & $          0.031$ \\
                                           $R_P/R_{\rm Jup}$& $          1.067$ & $          0.054$ \\
                                                 $R_P / R_S$& $          0.132$ & $          0.002$ \\
                                                   $i$~[deg]& $           88.1$ & $            0.5$ \\
                                                         $b$& $           0.39$ & $           0.09$ \\
                               $t_{\rm IV} - t_{\rm I}$~[hr]& $          2.743$ & $          0.033$ \\
                              $t_{\rm II} - t_{\rm I}$~[min]& $           22.2$ & $            2.0$ \\
                                            $T_c(363)$~[HJD]& $  2453787.70854$ & $        0.00035$ \\
                                            $T_c(366)$~[HJD]& $  2453799.75138$ & $        0.00032$ \\
                                                         $u$& $           0.49$ & $           0.05$
\enddata

\tablecomments{The parameter values in Column 2 are the median values
  $p_{\rm med}$ of the MCMC distributions. The quoted uncertainty is
  the average of $|p_{\rm med}-p_{\rm lo}|$ and $|p_{\rm med}-p_{\rm
    hi}|$, where $p_{\rm lo}$ and $p_{\rm hi}$ are the lower and upper
  68\% confidence limits. (The cumulative probability for values below
  $p_{\rm lo}$ is 16\%, and the cumulative probability for values
  above $p_{\rm hi}$ is also 16\%.)
  For the stellar mass, we use the value $M_S=0.81\pm 0.02$~$M_\odot$ from Santos et al.~(2006).
  The systematic errors due to the uncertainties in the stellar mass, the limb darkening function, and
  the reference flux are considerably smaller than the statistical
  errors and are not included here (see the text).}

\end{deluxetable}


\begin{thebibliography}

\bibitem[Agol et al.(2005)]{2005MNRAS.359..567A} Agol, E., Steffen,
  J., Sari, R., \& Clarkson, W.\ 2005, \mnras, 359, 567

\bibitem[Alard(2000)]{2000A&AS..144..363A} Alard, C.\ 2000, \aaps,
  144, 363

\bibitem[Alard \& Lupton(1998)]{1998ApJ...503..325A} Alard, C., \& Lupton, 
  R.~H.\ 1998, \apj, 503, 325

\bibitem[Baraffe et al.(2005)]{2005A&A...436L..47B} Baraffe, I.,
  Chabrier, G., Barman, T.~S., Selsis, F., Allard, F., \& Hauschildt,
  P.~H.\ 2005, \aap, 436, L47

\bibitem[Brown et al.(2001)]{2001ApJ...552..699B} Brown, T.~M.,
  Charbonneau, D., Gilliland, R.~L., Noyes, R.~W., \& Burrows, A.\
  2001, \apj, 552, 699

\bibitem[Charbonneau et al.(2002)]{2002ApJ...568..377C} Charbonneau,
  D., Brown, T.~M., Noyes, R.~W., \& Gilliland, R.~L.\ 2002, \apj,
  568, 377

\bibitem[Charbonneau et al.(2005)]{2005ApJ...626..523C} Charbonneau,
  D., et al.\ 2005, \apj, 626, 523
 
\bibitem[Claret(2000)]{2000A&A...363.1081C} Claret, A.\ 2000, \aap,
  363, 1081

\bibitem[Deming et al.(2005)]{2005Natur.434..740D} Deming, D., Seager,
  S., Richardson, L.~J., \& Harrington, J.\ 2005, \nat, 434, 740

\bibitem[Dobbs-Dixon et al.(2004)]{2004ApJ...610..464D} Dobbs-Dixon,
  I., Lin, D.~N.~C., \& Mardling, R.~A.\ 2004, \apj, 610, 464
 
\bibitem[Gaudi et al.(2005)]{2005ApJ...623..472G} Gaudi, B.~S.,
Seager, S., \& Mall\'en-Ornelas, G.\ 2005, \apj, 623, 472

\bibitem[Gelman \& Rubin(1992)]{Gelman.1992} Gelman, A.\ \& Rubin,
  D.~B.\ 1992, Stat. Sci., 7, 457

\bibitem[Gould et al.(2006)]{2006AcA....56....1G} Gould, A., Dorsher,
  S., Gaudi, B.~S., \& Udalski, A.\ 2006, Acta Astronomica, 56, 1

\bibitem[Holman \& Murray(2005)]{Holman.2005a} Holman, M.~J.\ \&
  Murray, N.~W.\ 2005, Science, 307, 1288
 
\bibitem[Holman et al.(2006)]{Holman.2006} Holman, M.~J., Winn, J.~N.,
  Latham, D.~W., O'Donovan, F.~T., Charbonneau, D., Bakos, G.,
  Esquerdo, G., Hergenrother, C., Everett, M.\ \& P\'al, A.\ 2006,
  ApJ, in press

\bibitem[Mandel \& Agol(2002)]{2002ApJ...580L.171M} Mandel, K., \&
  Agol, E.\ 2002, \apjl, 580, L171
 
\bibitem[Miralda-Escud{\'e}(2002)]{2002ApJ...564.1019M}
  Miralda-Escud{\'e}, J.\ 2002, \apj, 564, 1019
 
\bibitem[Pont et al.(2004)]{2004A&A...426L..15P} Pont, F., Bouchy, F.,
  Queloz, D., Santos, N.~C., Melo, C., Mayor, M., \& Udry, S.\ 2004,
  \aap, 426, L15

\bibitem[Pont et al.(2005)]{2005A&A...438.1123P} Pont, F., Bouchy, F.,
  Melo, C., Santos, N.~C., Mayor, M., Queloz, D., \& Udry, S.\ 2005,
  \aap, 438, 1123
 
\bibitem[Press et al.(1992)]{nr} Press, W.H., Teukolsky, S.A.,
  Vetterling, W.T., \& Flannery, B.P.\ 1992, Numerical Recipes in C
  (Cambridge: Cambridge Univ.\ Press)

\bibitem[Queloz et al.(2000)]{2000A&A...359L..13Q} Queloz, D.,
  Eggenberger, A., Mayor, M., Perrier, C., Beuzit, J.~L., Naef, D.,
  Sivan, J.~P., \& Udry, S.\ 2000, \aap, 359, L13
 
\bibitem[Rasio et al.(1996)]{1996ApJ...470.1187R} Rasio, F.~A., Tout,
  C.~A., Lubow, S.~H., \& Livio, M.\ 1996, \apj, 470, 1187
 
\bibitem[Rowe et al.(2006)]{2006ApJ...646.1241R} Rowe, J.~F., et al.\
  2006, \apj, 646, 1241

\bibitem[Santos et al.(2006)]{2006A&A...450..825S} Santos, N.~C., et
  al.\ 2006, \aap, 450, 825

\bibitem[Seager \& Mall{\'e}n-Ornelas(2003)]{2003ApJ...585.1038S}
  Seager, S., \& Mall{\'e}n-Ornelas, G.\ 2003, \apj, 585, 1038
 
\bibitem[Tegmark et al.(2004)]{2004PhRvD..69j3501T} Tegmark, M., et
  al.\ 2004, \prd, 69, 103501

\bibitem[Trilling(2000)]{Trilling.2000} Trilling, D.~E.\ 2000, \apjl,
   537, L61

\bibitem[Udalski et al.(2002)]{2002AcA....52..317U} Udalski, A., Szewczyk, 
  O., Zebrun, K., Pietrzynski, G., Szymanski, M., Kubiak, M.,
  Soszynski, I., \& Wyrzykowski, L.\ 2002, Acta Astronomica, 52, 317

\bibitem[Vidal-Madjar et al.(2003)]{2003Natur.422..143V} Vidal-Madjar,
  A., Lecavelier des Etangs, A., D{\'e}sert, J.-M., Ballester, G.~E.,
  Ferlet, R., H{\'e}brard, G., \& Mayor, M.\ 2003, \nat, 422, 143

\bibitem[Winn et al.(2005)]{2005ApJ...631.1215W} Winn, J.~N., et al.\
  2005, \apj, 631, 1215

\bibitem[Wolf et al.(2006)]{wolf06} Wolf, A.\ et al.\ 2006, ApJ, in
  press

\end{thebibliography}
\end{document}